\documentclass[10pt,letterpaper]{article}
\usepackage[top=0.85in,left=2.75in,footskip=0.75in]{geometry}

\usepackage{amsmath,amssymb}

\usepackage{changepage}

\usepackage{textcomp,marvosym}

\usepackage{cite}

\usepackage{nameref,hyperref}

\usepackage[right]{lineno}

\usepackage[nopatch=eqnum]{microtype}
\DisableLigatures[f]{encoding = *, family = * }

\usepackage[table]{xcolor}

\usepackage{array}

\newcolumntype{+}{!{\vrule width 2pt}}

\newlength\savedwidth

\usepackage{setspace} 
\raggedright
\usepackage[aboveskip=1pt,labelfont=bf,labelsep=period,justification=raggedright,singlelinecheck=off]{caption}

\makeatletter
\renewcommand{\@biblabel}[1]{\quad#1.}
\makeatother

\usepackage{lastpage,fancyhdr,graphicx}
\usepackage{epstopdf}
\fancyheadoffset[L]{2.25in}
\fancyfootoffset[L]{2.25in}
\begin{document}
\vspace*{0.2in}

\begin{flushleft}
{\Large
\textbf{Development of tools for all-cause mortality prediction in heart failure patients: a
systematic review} 
}
\newline
\\
Erik Aerts\textsuperscript{1,2* \Yinyang },
Annika Rosengren\textsuperscript{3,4 \ddag},
Martin Lindgren\textsuperscript{3,4 \ddag},
Martin Adiels\textsuperscript{5 \ddag},
Yinan Yu\textsuperscript{1,2,6 \Yinyang },
Helen Sjöland\textsuperscript{3,4,6 \Yinyang }
\\
\bigskip
\textbf{1} Computer Science and Engineering, Chalmers University of Technology, Gothenburg, Sweden
\\
\textbf{2} Department of Computer Science and Engineering, University of Gothenburg, Gothenburg, Sweden
\\
\textbf{3} Department of Molecular and Clinical Medicine, Sahlgrenska Academy, University of Gothenburg, Gothenburg, Sweden 
\\
\textbf{4} Department of Medicine, Geriatrics and Emergency Medicine, Sahlgrenska University Hospital, Gothenburg, Sweden
\\
\textbf{5} Institute of Medicine, Sahlgrenska Academy, University of Gothenburg, School of Public Health and Community Medicine, Gothenburg, Sweden
\\
\textbf{6} Center for Digital Health, Sahlgrenska University Hospital, Gothenburg, Sweden
\\
\bigskip

%
%
\Yinyang These authors contributed equally to this work.

\ddag These authors also contributed equally to this work.

\textcurrency Current Address: Dept/Program/Center, Institution Name, City, State, Country 

\dag Deceased

\textpilcrow Membership list can be found in the Acknowledgments section.

* aeerik@chalmers.se

\end{flushleft}


%
\section*{Abstract}
Heart failure is a serious medical condition associated with risk of acute deterioration, recurrent hospitalizations and high mortality making accurate mortality prediction potentially valuable for patient care and healthcare planning. Numerous prognostic models have been developed using conventional statistics (CS), machine learning (ML) and deep learning (DL), but variation in study populations, heart failure definitions, prediction horizons and development practices may limit interpretation of which models are most appropriate for use in different clinical settings. We conducted a systematic review of mortality prediction models published between 2009 to the start of 2026 searching three databases (PubMed, IEEE and Scopus) for studies predicting all-cause mortality over a 1–60 months prediction horizon among patients with heart failure. Data extracted were cohort characteristics, heart failure definitions, model development, cohort size, validation practices, mortality, computational model design and reported performance. The risk of bias and applicability were assessed using the Prediction Model Risk of Bias Assessment Tool (PROBAST). Of 1516 publications identified, 56 met the inclusion criteria. Publication numbers increased over the observation period, alongside increasing use of ML/DL models and decreasing use of CS approaches. Most development cohorts (64\%) were classified as \textit{Unknown/Undefined} heart failure, reflecting inconsistent reporting and terminology that did not correspond with established heart failure guidelines. External validation was reported in 23\% of studies. Reported model performance varied across prediction horizons, class imbalance and computational approaches. Overall, research in modeling heart failure mortality has grown alongside increasing adoption of ML/DL approaches, but considerable heterogeneity remains in cohort definition, model development, validation and reporting which limits utility. Greater standardization of study design and reporting is needed for meaningful comparison, reproducibility and evaluation of models for clinical application.


%
\section*{Author summary}
Heart failure is a common and serious condition that can lead to repeated hospitalizations and death. Predicting which patients are at higher risk could help support patient care and healthcare planning. Over the past two decades, many prediction models have been developed using both traditional statistical methods and newer machine learning approaches. We reviewed published heart failure mortality prediction models from 2009 to the start of 2026 to understand how the field has developed and how much variation exists between studies. We found that the number of prediction models increased over time, alongside growing use of machine learning methods. However, differences remained in how patients and heart failure cohorts were defined, how models were developed and how they were validated. A minority of studies reported external validation to showcase how well their model generalizes in new cohorts. Models using machine learning approaches showed slightly higher average reported performance for short range mortality prediction, although the studies differed substantially in number and design. Our findings highlight the need for more consistent study design, reporting and validation to make prediction models easier to compare and evaluate for clinical use.

\clearpage
\newgeometry{top=0.85in,left=1in,right=1in,footskip=0.75in}

\section*{Introduction}
Heart failure is a clinical syndrome characterized by the inability of the heart to pump blood at a rate meeting the functional demands of the body, often representing the end stage of structural or functional cardiac disease. The typical symptoms are dyspnea on exertion, fatigue, fluid retention, manifesting in swelling and edema. The severity of symptoms vary depending on heart failure progress \cite{schwinger-2021-pathophysiology-heart-failure}. In 2017 it was estimated to affect 64 million people worldwide and is expected to increase with an ageing population \cite{gbd-2017-disease-injury-incidence}. Heart failure represents a significant cause for hospitalization, commonly accompanied by other multiple long-term conditions. As heart failure progresses, periods of relative stability will often be punctuated by acute deterioration and hospitalizations in a partly unpredictable pattern, finally ending in death \cite{Shahim2023Global, jones-2017-prognosis-heart-failure}. Given the substantial clinical, human and health economic consequences of the condition, current prognostic assessments are insufficient and warrant improvement. 

 Efforts to improve heart failure prognostic modeling have commonly used the Seattle Heart Failure Model (SHFM), based on conventional statistics (CS) and published in 2006, as a benchmark \cite{levy-2006-seattle-heart-failure-model}. The model has been extensively validated on other cohorts with varying characteristics with respect to demographics and clinical severity. These validations have shown both higher- and lower discriminatory ability to identify high risk patients by SHFM compared to the originally published model
 \cite{hussain-2014-validation-seattle-heart-failure-model, shiraishi-2019-validation-seattle-heart-failure-model, shiraishi2017validation, li2019validity, regoli2013validation}. At present, the SHFM is the most widely used prediction tool for heart failure patients \cite{BILCHICK2020924}. Since its publication, there has been considerable advancements in data collection, management, computational capacity, understanding of prognostic clinical factors and new modeling techniques have been introduced. This has opened the door for more complex and refined machine learning (ML) and deep learning (DL) models to be developed to improve prognostic predictions, resulting in a multitude of prognostic tools targeted to clinical use. Previous reviews have compared the performance in areas such as forecasting of mortality and/or re-hospitalisations for heart failure patients within a given time frame of CS and ML models, generally tending to favor the latter \cite{Shin2021}. However, understanding how heart failure prognostic modeling has developed over time and the extent of variation in the development is important for interpreting these reported performance differences to understand which models stands out from the masses. 
 
 To address this, this systematic review addresses temporal trends in heart failure mortality prediction modeling, together with the decisions underlying model design and validation. We characterize the variation between studies in patient cohort composition and their respective heart failure definitions, validation strategies and modeling approaches, establishing the landscape within which reported model performance can be interpreted. Reported performance is subsequently examined across subgroups defined by heart failure severity as reflected by outcome prevalence, prediction horizon and computational model design. These subgroups are used to describe reported performance across different modeling settings for CS and ML/DL approaches.

\section*{Methods}

\subsection*{Search strategy and selection}
 This systematic review was conducted in agreement with the Preferred Reporting Items for Systematic Reviews and Meta-Analyses (PRISMA) \cite{Page2021}, by performance of a literature search in the databases PubMed, IEEE and Scopus for articles published between January 1st 2009 to January 1st 2026 to study published research from contemporary medical practice. The search terms included: \textit{mortality}, \textit{model}, \textit{heart failure}, \textit{artificial intelligence}, \textit{machine learning}, \textit{deep learning} which were used to create search strings in the databases. A complete list of search strings used for the literary search can be found in Table 1 in S1 Appendix. One author (E.A.) performed the selection process, starting with screening titles and abstracts based on topical relevance. Selected abstracts were manually filtered for relevance and a subset of articles was selected for a full-text review. Duplicates were removed. The inclusion criteria were: Prediction models for all-cause mortality in heart failure. Exclusion criteria from full-paper review were: i) no use of statistical methods to develop the model or validation of an existing model on a new cohort, (ii) time frame of predictions outside a preselected range of 1-60 months, (iii) outcome included in-hospital mortality during the initial hospitalization at the start of the study or outcome combined hospitalization with mortality (iv) cohorts derived from patients with heart failure who underwent a surgical intervention  (v) inaccessible paper or not published in English, (vi) missing information during data collection. No minimum development cohort size threshold was imposed to preserve a comprehensive representation of the published literature. No protocol was prospectively registered or published before the review was conducted.

\subsection*{Data collection}

\noindent One of the authors (E.A) performed the data collection process for each paper, which included (i) name- and year of publication, (ii) class- and technique of the model, (iii) size of the development cohort, (iv) class imbalance ratio from the cohort mortality rate, (v) the estimated complexity group of the model, (vi) mortality prediction range, (vii) model performance and evaluation metric, (viii) performance of external validation (yes/no), (ix) type of heart failure. Each model was placed in the CS, ML or DL class based on the provided description of the model. Each model was further categorized based on the modeling technique used in development with a detailed list of the categories and their respective descriptions provided in Table 2 in S1 Appendix. If multiple modeling techniques were used, the model technique was annotated as \textit{Ensemble model}. Logistic regression was classified according to how it was used and described in the original study. When used as a conventional statistical modeling approach, it was classified as CS. When presented as a machine learning approach within a predictive modeling framework, it was classified as ML. 

We adopted the definitions established by the European Society of Cardiology (ESC) to create four groups, as outlined in the \textit{2021 ESC Guidelines for the diagnosis and treatment of acute and chronic heart failure} in this report \cite{ESC}, using the heart failure categories defined in the ESC guidelines (HFrEF: reduced ejection fraction with signs/symptoms of heart failure and evidence of cardiac dysfunction (left ventricular ejection fraction (LVEF) $\leq$40\%); HFmrEF: mildly reduced ejection fraction with signs/symptoms of heart failure and evidence of cardiac dysfunction (LVEF 41–49\%); HFpEF: preserved ejection fraction and evidence of cardiac dysfunction (LVEF $\geq$ 50\%) plus objective evidence of cardiac structural and/or functional abnormalities consistent with left ventricular diastolic dysfunction or raised filling pressures) \cite{ESC}. Each study cohort was assigned to one of the four heart failure groups. Heart failure groups and their descriptions can be found in Table 3 in S1 Appendix. External validation was defined as performed in a cohort derived from an external hospital or patient group. If the validation was performed on unseen data in the same dataset or a temporal shift from the same patient cohort, the validation was deemed as internal. For missing data, or unclear data leading to missing data, for any of the features during the data collection process the paper was excluded. However, when heart failure description did not conform with standard definitions as above, was mixed or not described, the type of heart failure was annotated as "\textit{Not specified}" rather than being excluded. 

Different models were subgrouped based on interpretability. We separated the CS models associated with high interpretability in the decision making process from the ML/DL models where the decision making process is typically less transparent and often requires additional methods to be fully understood. We studied differences in model performance between the two groups for different time ranges of mortality prediction and class imbalance in the datasets. The mortality prediction ranges were binned into three groups: short range prediction (1-6 months), medium range prediction (6$<$ to $<$18 months) and long range prediction (18-60 months). Class imbalances were grouped as: high imbalance ($<$15\% or $>$85\% mortality), medium imbalance (15\%-25\% or 75\%-85\% mortality) and low imbalance ($>$25\% and $<$75\% mortality). Moreover, we explored differences between models based on computational demands in development and usage. Determining if high computational resources (e.g. high-performance graphics processing units (GPUs) or large memory capacity) were likely needed for model development was concluded from the model description made in the article. The computational requirements for training and deployment was grouped as: Group 1) Models assumed to not require high-resource computational infrastructure to train- or run the model, primarily consisting of traditional CS and lower-complexity ML approaches that did not explicitly state the need for advanced computational infrastructure; and Group 2) Models assumed to require high-resource computational infrastructure for training- and running the model, mainly DL models, ensemble-based ML approaches, or studies explicitly reporting the use of computationally intensive training procedures or high-performance computing infrastructure. The recorded performance metrics were the \textit{C}-statistic and the area under the receiver operating characteristic curve (AUC). The \textit{C}-statistic was selected as the primary metric due to its widespread use in prognostic modeling and frequent reporting across the included studies, while AUC was recorded as a secondary metric. Although the \textit{C}-statistic and AUC can be equivalent for binary outcomes and fixed timeframes, their relationship differs for time-to-event outcomes depending on the definition and estimation method used \cite{cstatistics}. Therefore, both metrics were retained when reported together to accurately reflect the performance measures reported in the original studies. As model performance from development data may cause overfitting, the study prioritized the model performance from a validation set if available. If multiple model architectures were tested in the same paper, the final model with the highest resulting performance was selected for our study. If varying model architectures performed better at different time intervals, the study selected the highest performing model per time interval. Continuous variables are reported as mean ($\mu$) and standard deviation ($\sigma$) and categorical values as proportions. Descriptive comparisons were used to examine patterns across modeling groups and study characteristics. Where statistical significance between distributions was assessed, the Mann-Whitney U test was used.

\subsection*{Quality assessment}
Each paper was independently assessed by two of the authors (E.A., M.L., H.S., Y.Y.) using the PROBAST tool, which evaluates risk of bias and applicability in prediction models. Concerns were graded as Low (\textit{L}), Unclear (\textit{U}), or High (\textit{H}) across four domains: Participants, Predictors, Outcome and Analysis. Scoring was based strictly on information reported in the original publications and therefore reflects both methodological quality and completeness of reporting, as required by the PROBAST framework. A joint overall assessment of concern in risk of bias- and applicability was graded using the same metric system. For averages, the grades were converted into an ordinal scale as follows: \textit{L}=1, \textit{U}=2, \textit{H}=3. Reviewers included clinical cardiologists ($n$=2) and data scientists ($n$=2).

\section*{Results}
The initial literary search resulted in a total of 1516 articles. After initial screening of the titles, 31 duplicates were identified, 933 articles were deemed as out of scope, and 552 articles were retained for abstract screening. We conducted a full-text review from 90 of the selected in scope abstracts and identified a total of 56 articles fulfilling the requirements for the final set \cite{nr1Guo2024,nr2Subramanian2011,nr3Richter2013,nr4Camara, nr5Carluccio,nr8Voors2017,nr9Escobar2017,nr10Karauzum2018,nr11Bowen2018,nr12Salah2014,nr13WANG2020103340,nr14Hwang2020, nr15Suzuki2020, nr18Lin2022,nr20Zhuang2024, nr21Ketabi2024, nr23Valsaraj2023, nr24Wang2018, nr25wang2020multi, nr26cai2022machine, nr27adler2020improving, nr28kwon2019artificial, nr29WANG2020105383, nr32WANG2021104813, nr38angraal2020machine, nr39beecy2020utilizing, nr40ju2021derivation, nr41negassa2021prediction, nr42tohyama2021machine, nr43takabayashi2021clinical, nr44vandergalien2021prediction, nr46park2022predicting, nr47herman2022utilizing, nr48diagnostics12122947, nr55takahama2023clinical, nr56tian2023machine, Sabouri2023, nr59xu2023interpretable, nr60benassuli2024expert, nr66zhang2024predicting, nr69garcia2020short, nr74article, nr75magaldi2024predicting, nr76simpson2020prognostic, Li2023, Chadwick2025, Wang2025, Gao2025, Wu2025, Scrutinio2025, Shin2025, Lee2025, Weller2025, Atias2025, Alahdab2025, Aoki2025}. The PRISMA flow chart is presented (Fig 1). Recorded features for each article are presented in Table 4 in S1 Appendix. 

\begin{figure}
    \centering
    \includegraphics[width=0.5\linewidth]{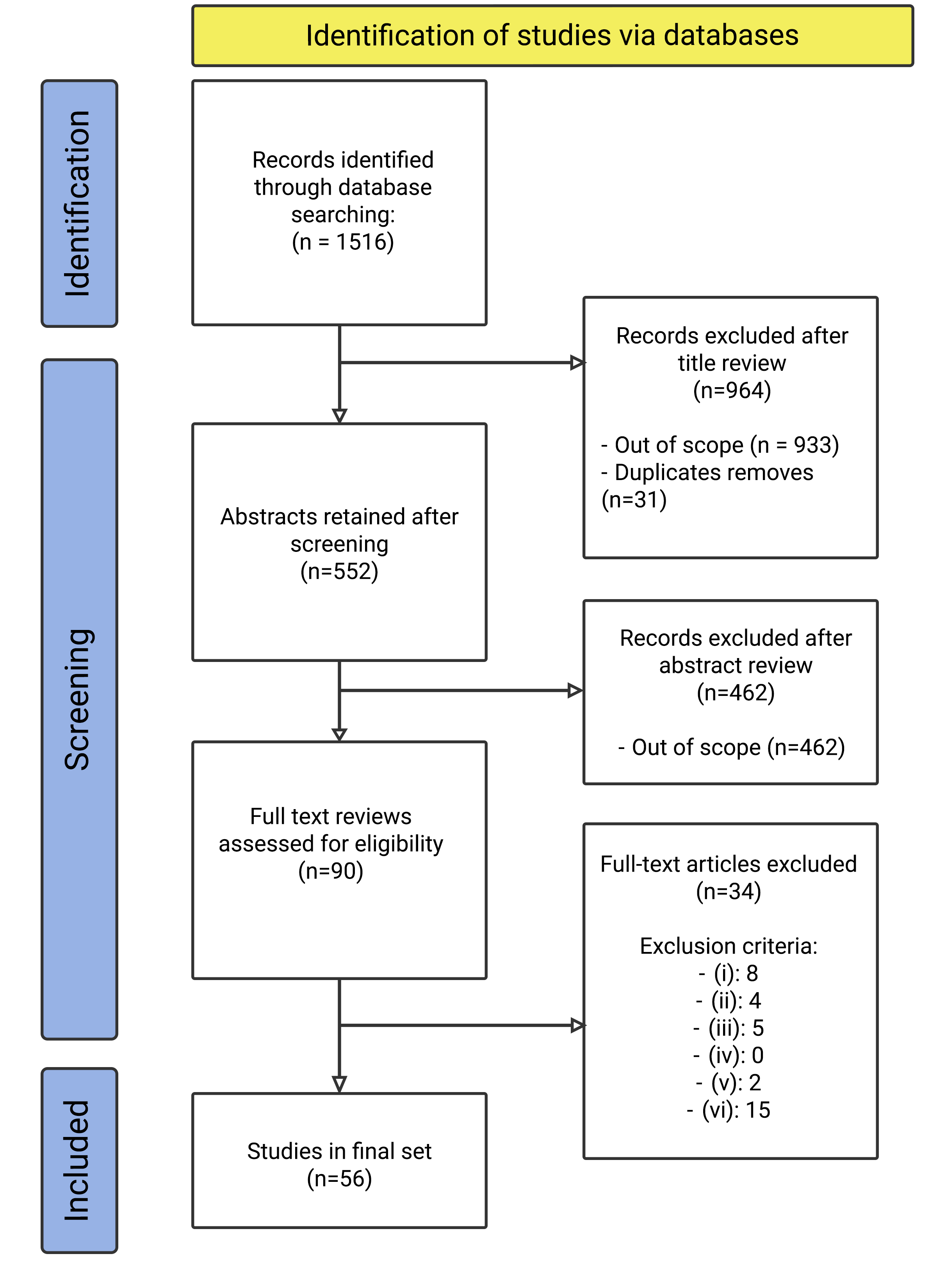}
    \caption{\textbf{Preferred Reporting Items for Systematic Reviews and Meta-Analyses (PRISMA) flow chart.}}
    \label{fig:PRISMA}
\end{figure}
 
 In total, $n$=13 articles (23\%) used external validation. Performance was reported with \textit{C}-statistics in $n$=22 articles (39\%), AUC in $n$=32 articles (57\%) and with both metrics in $n$=2 articles (4\%). The average cohort size was $\mu$=4153 ($\sigma$=5787; range 117–35732), generally larger for ML/DL ($\mu$=4632, $\sigma$=4611) compared to CS ($\mu$=3220, $\sigma$=7486) models ($p$ $<$ 0.01). A majority of articles ($n$=48) used specific clinical terms to define their heart failure cohort with $n$=18 different terms used. In total, $n$=36 (64\%) were classified in the heart failure group "\textit{Unknown/Undefined}" followed by $n$=12 (21\%) "\textit{HFrEF+HFmrEF}", $n$=5 (9\%) "\textit{HFpEF}" and $n$=3 (5\%) "\textit{Ischemic/CHD}". The number of publications per year increased from 2009 to the start of 2026. The proportion of articles related to ML/DL models increased over the observation time frame while the proportion of articles using CS decreased (Fig 2). The modeling techniques used were: for CS ($n$=21), distributed as Cox (57\%) or logistic regression (43\%) models. For ML/DL ($n$=35), a total of 10 different techniques were used, most frequently Boosted decision tree algorithms (40\%), followed by logistic regression (17\%) and ensemble models, convolutional neural networks, kernel-based methods and random forest (all 9\% each). 

 \begin{figure}
    \centering
    \includegraphics[width=0.85\linewidth]{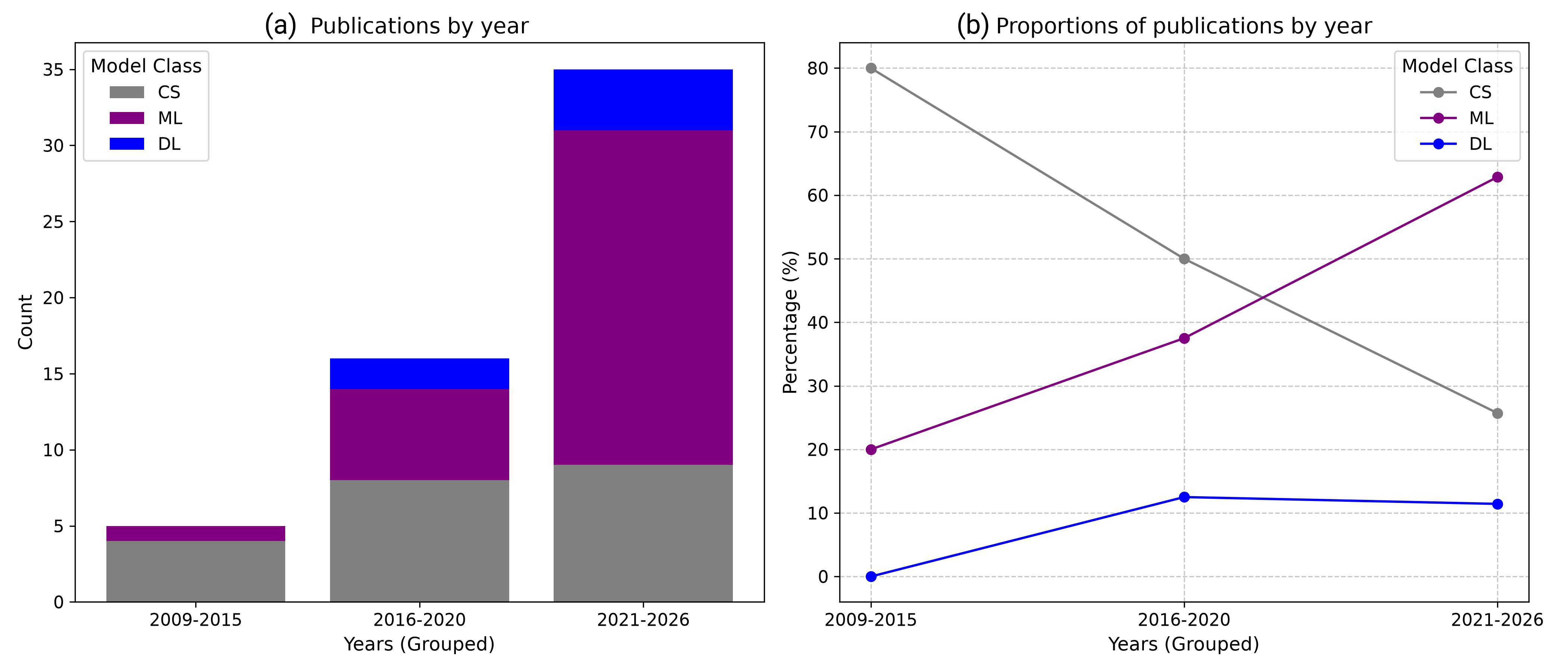}
    \caption{\textbf{(a): Bar diagram of the number of publications in the study by grouped year ranges. (b): Line diagram of the percentages of publications related to a model class for the grouped year ranges. } The colors of the bars and lines are consistent between subfigures. The color of the bars for each year range indicated the proportion of publications related to a model class.\\
    \normalfont\textit{Abbreviations: CS = Conventional statistics; ML = Machine learning; DL = Deep Learning}}
    \label{fig:over time}
\end{figure}
 
 Reported performance in \textit{C}-statistic and AUC were stratified by dataset imbalance groupings and visualized using box plots for descriptive comparison between ML/DL and CS studies. Only the primary metric in \textit{C}-statistic was included in the comparison if an article reported both metrics. High class imbalance was more common in ML/DL models (60\%) than medium- and low class imbalance (20\% each). In contrast, low class imbalance was more common for CS (42\%) in comparison to medium- and high class imbalance (29\% each) On average, high- and low class imbalances had high box overlapping between ML/DL and CS models, whereas medium class imbalance showed less box overlap (Fig 3). Overall, the reported performance of the models showed similar distributions across the three class imbalance groups. The models included in the review were applied to a variety of prediction time ranges. Short range mortality prediction models were more common for ML/DL (43\%) than medium- and long range mortality prediction (34\% and 23\% respectively). Among CS models, medium range mortality prediction was the most common (48\%), followed by long range (43\%) and short range (9\%) prediction tasks. Similarly, reported \textit{C}-statistics and AUC were grouped by prediction time ranges and visualized using box plots (Fig 4). For short range mortality prediction, the distributions of reported performance showed a higher average for ML/DL models than for CS models, with no overlap between the heterogeneous distributions. The number of studies differed substantially between the two groups, with short range mortality being the most frequently studied time range among ML/DL models but the least frequently studied among CS models. Reported performance showed similar tendencies and frequencies in the other two time ranges, with great box overlap between models for the medium range and long range mortality prediction. The majority of models were of lower complexity (Group 1 consisted of 86\% of studies, Group 2 of 14\% of studies). Reported performance distributions were broadly similar between the two groups, while the more resource-intensive models showed a higher average and less variation in reported performance (Fig 5).

 \begin{figure}
    \centering
    \includegraphics[width=0.7\linewidth]{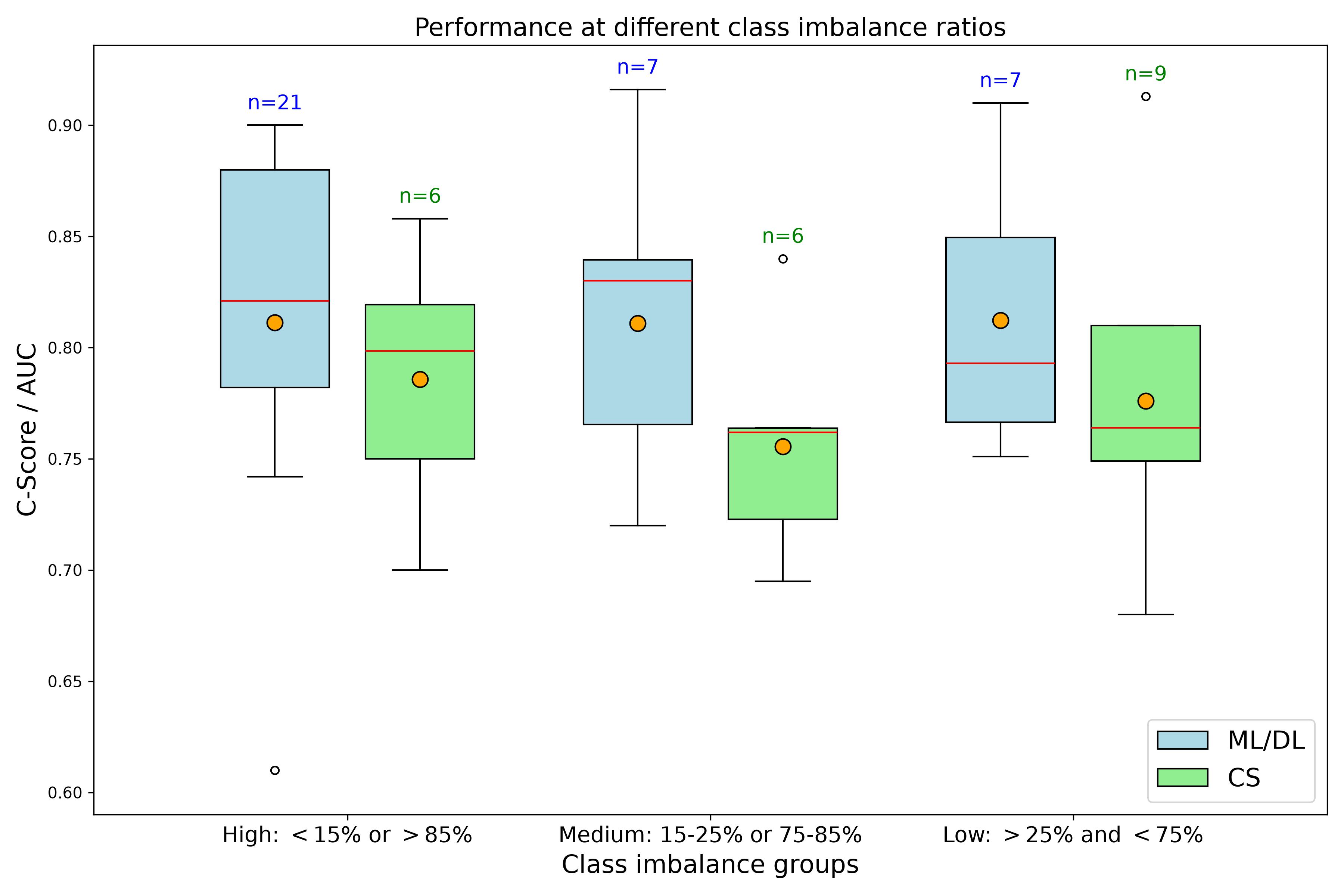}
    \caption{\textbf{Boxplot of the reported performance for the developed models grouped by dataset class imbalance.} The blue boxes represents the performance distribution for ML/DL models and the green boxes represents the performance distribution for CS models for each class imbalance range.  The median of the distribution is presented as a red line on each box. The mean of the distribution is presented as a orange dot. \\
    \normalfont\textit{Abbreviations: CS = Conventional statistics; ML = Machine learning; DL = Deep Learning}}
    \label{fig:class imbalance}
\end{figure}

\begin{figure}
    \centering
    \includegraphics[width=0.7\linewidth]{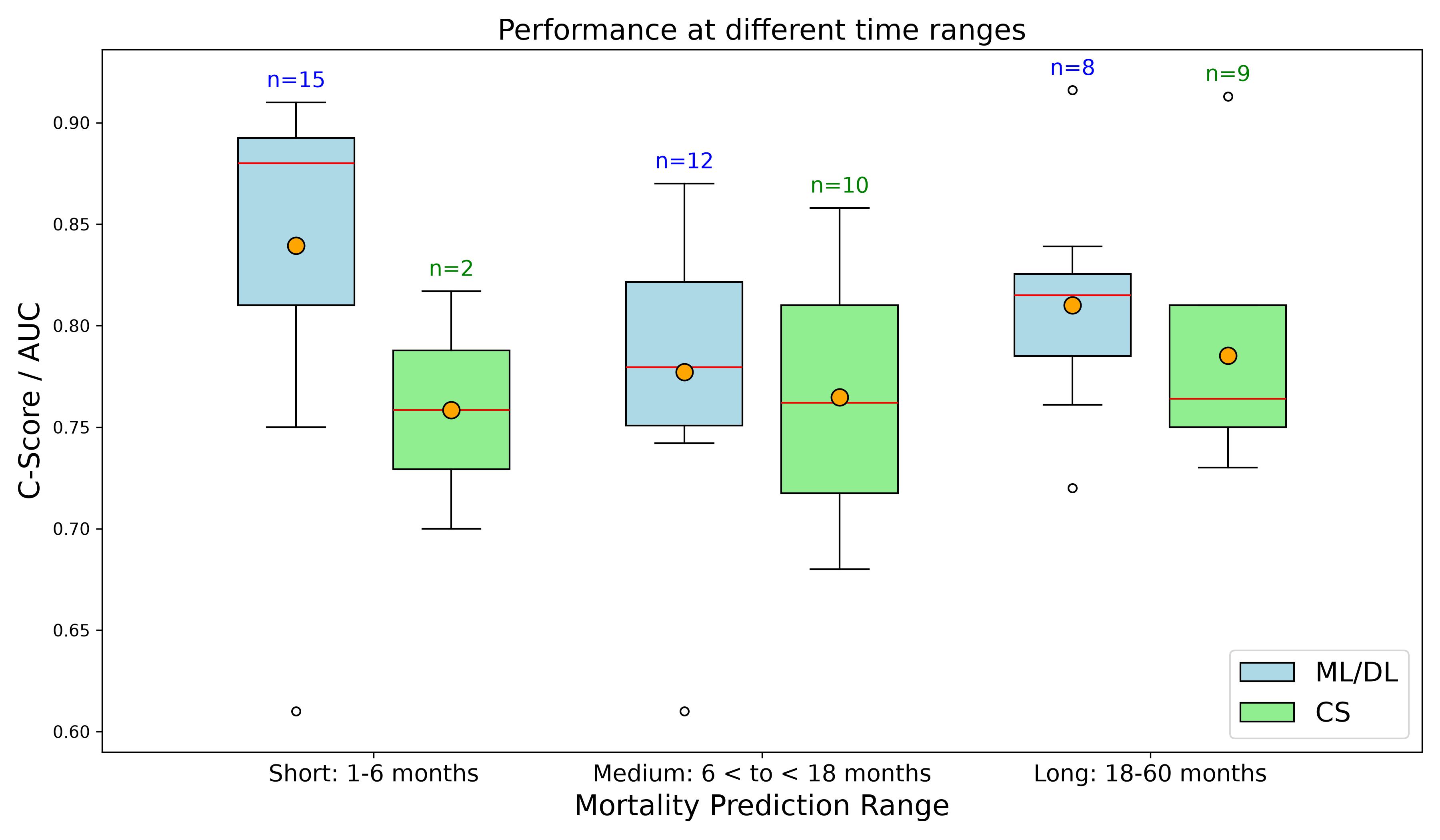}
    \caption{\textbf{Boxplot of reported performance for the developed models grouped by mortality prediction range.} The blue boxes represents the performance distribution for ML/DL models and the green boxes represent the performance distribution for CS models for each time range.  The median of the distribution is presented as a red line on each box. The mean of the distribution is presented as a orange dot.  \\
    \normalfont\textit{Abbreviations: CS = Conventional statistics; ML = Machine learning; DL = Deep Learning}}
    \label{fig:timeranges}
\end{figure}

\begin{figure}
    \centering
    \includegraphics[width=0.7\linewidth]{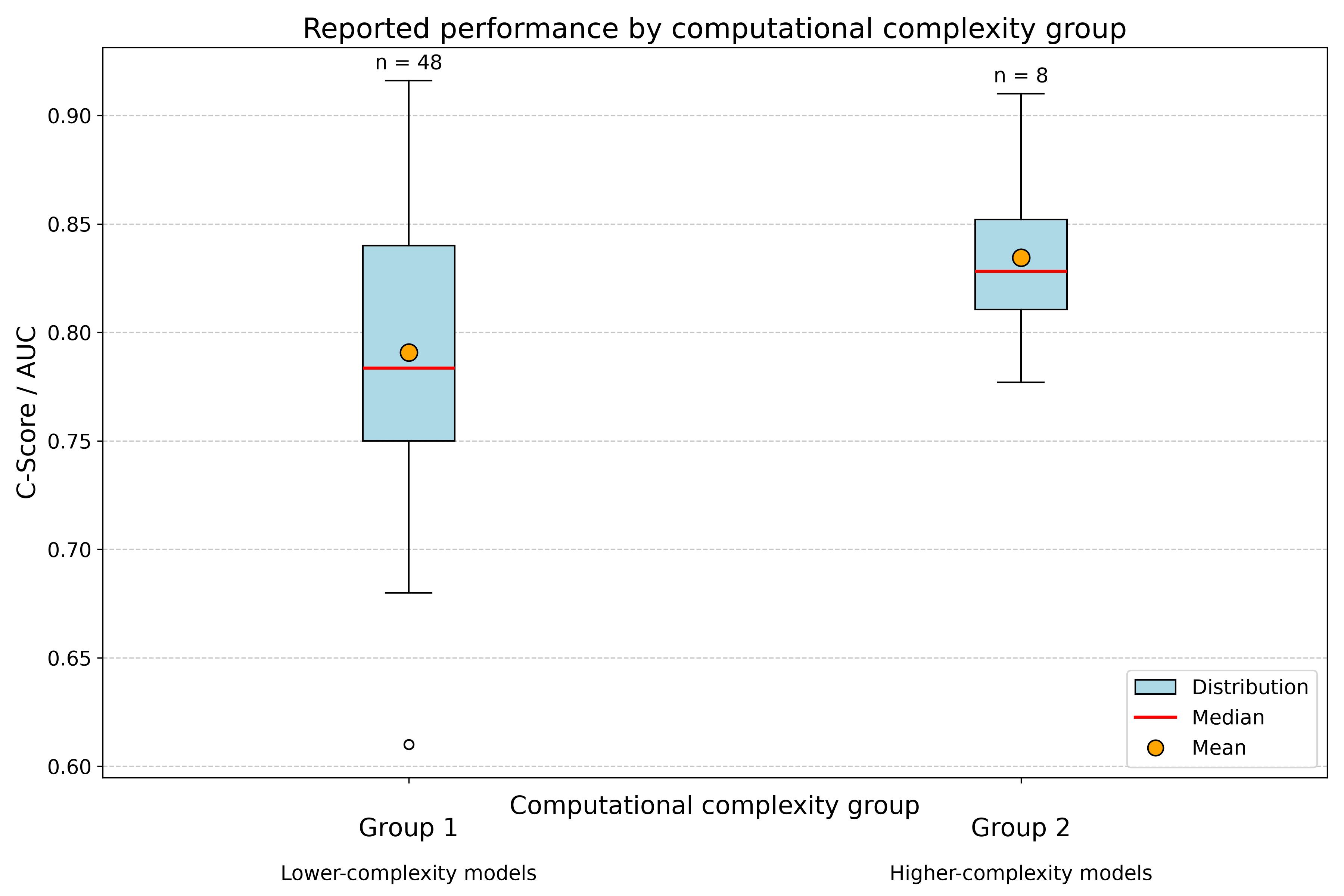}
    \caption{\textbf{Boxplot of the reported performance for the developed models grouped by estimated complexity group.} The median of the distribution is presented as a red line on each box. The mean of the distribution is presented as a orange dot. }
    \label{fig:complexity}
\end{figure}

\subsection*{Quality assessment}

The assessments of risk of bias and applicability for each model can be found in Table 5 in S1 Appendix. Using the PROBAST tool, the majority of articles showed low or intermediate concern for bias (\textit{L}: 37.5\%, \textit{U}: 37.5\%, \textit{H}: 25\%) with an average score of 1.88. Bias scored similarly for CS models (1.90) and ML/DL models (1.84) and was not significant ($p=$0.874). For applicability in all articles, a low to intermediate concern was found (\textit{L}: 46\%, \textit{U}: 29\%, \textit{H}: 25\%), with an average score of 1.82. CS models (1.86)  and ML/DL (1.80) models scored similarly in concern of applicability and the difference was not significant ($p=$0.950).

\section*{Discussion}
In this systematic review of prediction models for heart failure mortality, we observed considerable variation in model development and evaluation, with reported performance examined across different class imbalance groups and mortality prediction ranges. The cohort setups for the studied articles were diverse. First, considerable variation was observed in the number of patients in the cohorts, the type of heart failure and the reported mortality. The majority of the terminology used to describe patient populations was not consistent with established clinical classification standards for heart failure, which reduced comparability between studies \cite{ESC}. These inconsistencies may reflect the inclusion of older studies, use of heterogeneous or outdated definitions, or the absence of clear specification in the original publications resulting in many cohorts being categorized as \textit{Unknown/Undefined}. The diversity in data definitions and cohort characterization creates uncertainty regarding comparison between models highlighting the importance of adopting internationally established heart failure classification systems in future studies to improve comparability and generalizability. Moreover, the proportion of studies employing external validation to verify model performance was low, limiting confidence in the generalizability of reported results. Without standardized cohort definitions and consistent validation procedures, it becomes difficult to determine whether improved performance reflects genuine model innovation or merely overfitting to narrowly defined datasets. This heterogeneity also impedes systematic comparisons and weakens the foundation for clinical adoption. Therefore, establishing common reporting standards for cohort composition, heart failure classification and validation practices preferably aligned with current guidelines would be a critical step toward improving model transparency, reproducibility and clinical relevance. \\
\indent Between 2009 and the start of 2026 the use of CS models decreased in favor of ML models, reflecting the increased accessibility of ML modeling during this period. Well documented ML packages such as Tensorflow, scikit-learn and xgboost for common programming languages like Python and R released stable versions around 2011–2016, as acknowledged by the ML based articles in the study \cite{Abadi2016, Pedregosa2011, Chen2016}.  These packages lowered the threshold of skill needed in programming and ML mathematics to adapt these architectures. A similar trend may be followed to develop future DL models, as more sophisticated programming packages get released. However, the large quantity of data needed to train the DL models and the additional computational resources may restrict this rise even with increased programming convenience. Among CS models in the review, techniques were used evenly. In contrast, boosted decision trees dominated ML based studies, appearing in about a third of the studied ML articles. Missing values is a common and expected feature of real-world clinical data which may cause challenges for ML models. Boosted decision tree models can be advantageous when working with missing predictor values. For example, XGBoost, a commonly used boosted tree algorithm, can handle missing values during model development and prediction without requiring prior imputation, whereas other modeling approaches may be restricted by requiring all predictor values to be available \cite{Rezvan2015, Chen2016}. ML models are often described as black-box systems, where understanding the basis of individual predictions can be difficult \cite{Gao2022}. Tree-based models can provide some additional insight to feature importance measures derived from the fitted trees. However, these measures do not necessarily provide a complete explanation of individual predictions \cite{Agarwal2022}. Such information may be useful in clinical settings where understanding the factors contributing to model predictions is important.\\
\subsection*{Implications and future directions}
\noindent We observed a general trend of increased tool development over the studied time period. However, a larger number of prediction models does not necessarily translate into clinical utility. Comparison of tools to determine the most appropriate model for a specific patient population, outcome, prediction horizon, or clinical setting requires considerable effort, which is likely to increase as the number of available models continues to grow. Similar concerns have been described in other areas of medicine, where the proliferation of outcome assessment tools has created difficulties in selecting appropriate instruments and has highlighted the need for guidance to support tool selection and comparability \cite{Harding2011}. In the context of heart failure mortality prediction, the heterogeneity identified in this review suggests that model selection may become increasingly complex as the number and diversity of available prediction tools continues to grow creating a tool fatigue problem. 

 Large language models (LLMs) have the potential to assist clinicians in the selection of appropriate prediction tools. LLMs have demonstrated promising capabilities in tool learning, where a model selects and invokes an appropriate tool from a set of available tools to complete a given task. A key component of this process is tool retrieval and selection, in which the LLM identifies candidate tools and determines which are appropriate for a given task based on their descriptions and the requirements. Foundational approaches such as Toolformer and ReAct demonstrated the potential for LLMs to interact with external tools and incorporate their outputs into a broader reasoning process \cite{Toolformer2023, Yao2023React}. Subsequent work has examined tool selection as a distinct component of tool use. For example, MetaTool introduced a benchmark specifically designed to evaluate whether LLMs can recognize when tools are needed and select appropriate alternatives from a set of candidates, including scenarios involving similar tools, different contexts and multiple tools \cite{MetaTool}. Subsequent work has extended these capabilities to medical applications. For example, MeNTi introduced an agent framework for medical calculators that supports flexible tool selection and nested tool calling, enabling LLMs to identify and apply appropriate calculators to clinical questions \cite{MeNTi2025}. More recently, AgentMD extended this approach specifically to clinical risk prediction by retrieving candidate risk calculators and using an LLM to select appropriate tools based on their applicability to the patient and clinical context, before applying the selected calculators and interpreting their results \cite{Jin2025AgentMD}. Together, these studies demonstrate the potential for LLMs to act as an intermediary layer between clinical questions and existing computational tools, rather than generating predictions directly. A similar approach could potentially be applied to the selection of clinical prediction models. Given a clinical question and information about the target population, an LLM could compare these characteristics with the reported development and validation characteristics of available prediction tools and identify models appropriate for the intended application. Such an approach could help clinicians navigate the landscape of prediction models and reduce the need to develop yet another model when a suitable tool already exists. Highlighting this role for LLMs in clinical practice is important to distinguish their potential as a tool-selection and retrieval layer from their use as a direct source of clinical prediction. Recent evidence suggests that LLMs are not yet ready for deployment as standalone medical assistants in direct patient care, despite promising performance on established benchmarks and clinical scenarios \cite{Bean2026}. At the time of writing, LLMs may therefore be more appropriately used to support the identification and interpretation of existing validated prediction models, while clinical risk assessment remains based on models developed and validated for the relevant population and outcome. 
 \subsection*{Quality assessment and limitations}
\noindent The quality assessment with PROBAST showed broadly similar levels of concern for risk of bias and applicability across CS and ML/DL models. The PROBAST assessment of risk of bias includes methodological and technical details of model development. Such information may be reported differently across modeling approaches, particularly where ML/DL studies which usually provide more extensive descriptions of computational procedures. Applicability is assessed based on the study population, predictors and outcome timing while aspects such as model complexity and interpretability are not considered within the framework. Therefore, PROBAST provides an assessment applicability based on reported study characteristics, but does not capture all aspects relevant to comparing different modeling approaches.\\
\indent Our study had a few limitations. First, heterogeneity in cohort size, the use of external validation and heart failure definitions, may affect the comparability of reported performance estimates and restrict the generalizability of the findings. Second, assigning studies to predefined groups for heart failure and model complexity required interpretation of the descriptions provided by the original authors, particularly when reported characteristics did not clearly correspond to a predefined category. Third, the review protocol was not prospectively preregistered in a public registry prior which may increase the risk of reporting bias despite the use of predefined inclusion and exclusion criteria. These factors may restrict the comprehensiveness of our review and should be considered when interpreting the relative performance of ML/DL and CS models.

\section*{Conclusions}

In conclusion, heart failure mortality prediction modeling represents a heterogeneous field with substantial variation in cohort definitions, modeling approaches, development and validation practices. External validation was reported in only a minority of studies. ML/DL models showed higher average reported performance than CS models for short range mortality prediction, although this result should be interpreted considering differences in study frequency and developmental circumstances. For clinical usability and comparative assessment of model performance, future studies would benefit from standardized cohort definitions based on international guidelines and from external validation.  
\nolinenumbers

\section*{Supporting information}
\textbf{S1 File Appendix. Supporting tables for search strings, grouping definitions and extracted data from articles.} The tables are provided as a PDF file named \texttt{S1\_file.pdf}.

\section*{Author contributions}
\textbf{Conceptualization}: Erik Aerts, Yinan Yu, Helen Sjöland\\
\noindent \textbf{Data Curation}:  Erik Aerts\\
\noindent \textbf{Formal Analysis}: Erik Aerts \\
\noindent \textbf{Funding Acquisition}: Yinan Yu, Helen Sjöland \\
\noindent \textbf{Investigation}: Erik Aerts, Yinan Yu, Helen Sjöland, Martin Lindgren \\
\noindent \textbf{Methodology}: Erik Aerts, Yinan Yu, Helen Sjöland \\
\noindent \textbf{Project Administration}: Erik Aerts \\
\noindent \textbf{Supervision}: Yinan Yu, Helen Sjöland \\
\noindent \textbf{Validation}: Yinan Yu, Helen Sjöland \\
\noindent \textbf{Visualization}: Erik Aerts \\
\noindent \textbf{Writing – Original Draft Preparation}: Erik Aerts, Yinan Yu, Helen Sjöland \\
\noindent \textbf{Writing – Review \& Editing}: Martin Lindgren, Martin Adiels, Annika Rosengren

%
%

\bibliography{plos_bibtex_sample}

\end{document}